# Influences of Si sheet doping densities on the morphological, conductive and optical characteristics of InAs/GaAs quantum dots

Ke-Fan Wang, X. G. Yang, Y. X. Gu, H. M. Ji, T. Yang[a] and Z. G. Wang

*Key Laboratory of Semiconductor Materials Science, Institute of Semiconductors, Chinese Academy of Sciences, P.O. Box 912, Beijing 100083, China*

The influences of Si sheet doping levels on the properties of InAs/GaAs quantum dots (QDs) are investigated by atomic force microscopy (AFM) and photoluminescence (PL). AFM measurements reveal that Si sheet doping doesn't change the morphology of InAs QDs. Conductive AFM exhibits a quick current decrease when the Si doping density reaches $5 \times 10^{11} cm^{-2}$. PL measurements show that the Si doping can significantly enhance the PL intensity. The PL peak intensity of InAs QDs doped to $5 \times 10^{11} cm^{-2}$ is increased about thirty-five times from that of the undoped ones at 300K. The results observed here can be explained by a supposed positive-charged, strain-relaxed Si-doped thin InAs layer inside the InAs QDs.

---

[a] Electronic mail: tyang@semi.ac.cn.

Self-assembled InAs/GaAs quantum dots (QDs) via Stranski-Krastanow growth mode have attracted considerable attention due to their atomic like density of states, and excellent optical properties, as well as for potential applications in various optoelectronic devices, such as QD lasers, optical amplifiers, etc.[1,2] Recently, Inoue *et al.*[3] found that if InAs/GaAs QDs were doped with Si during the assembling and self-limiting growth stages, the QDs would have a much stronger PL intensity than those undoped or doped at nucleation stage. This finding is interesting, but contradictory to the previous reports that Si doping will degrade the optical quality of InAs QDs.[4,5] In this letter we report on the influences of Si doping densities on the morphological, conductive and optical properties of the QDs.

InAs QDs samples were grown on $n^+$-GaAs(001) substrates in a Vecco Gen-II molecular beam epitaxy system. First, a 300-nm-thick GaAs buffer layer was deposited on a GaAs substrate at a rate of 0.49 monolayers (MLs)/s at 600℃. Subsequently, 2.0-MLs of InAs were deposited at a rate of 0.02 MLs/s at 500℃. After this step, the *in situ* reflection high energy electron beam diffraction showed a transition from elongated streaks to ordered spots that indicated the initial formation of InAs QDs, corresponding to the onset of the self-limiting growth stage.[6] After a one minute interruption, 0.6-MLs InAs was deposited for 30 seconds to finish the growth. During the previous step, Si atoms with different sheet doping levels, (4, 10, 50, 100)×$10^{10}$cm$^{-2}$, were co-deposited with In atoms during the first 12 seconds. The Si doping levels were calculated based on Hall measurements. The QDs were capped with a 10 nm GaAs layer deposited at 500℃, followed by a 40 nm GaAs deposited at 600℃. Finally, on the GaAs cap layer surface, the growth sequence only of the InAs QDs was repeated. The surface InAs QDs' morphology and current distributions were obtained simultaneously by a conductive AFM system (Seiko Instrument, SPA-300HV) in contact mode using a rhodium-coated silicon tip. To obtain more accurate morphology, AFM measurements were repeated in tapping mode by another system (Seiko Instrument, E-sweep). The PL spectrum of the InAs QDs was measured by using a solid state laser (532nm), an iHR320 spectrometer (Jobin Yvon) and a liquid-nitrogen-cooled InGaAs detector. The QDs sample was cooled by a closed-cycle helium cryostat and could range from 13K to 300K.

Fig. 1 shows the morphological images of the surface InAs QDs with different doping levels: (a) undoped; (b) $1\times10^{11} cm^{-2}$; (c) $5\times10^{11} cm^{-2}$; (d) $1\times10^{12} cm^{-2}$. Consistent with the previous report,[3] the morphologies of the InAs QDs with different Si doping have no obvious changes compared to that of the undoped InAs QDs, and all of them have mean diameters of 35~37 nm, mean heights of 7~8 nm, densities of 3~4$\times10^{10}$ cm$^{-2}$ and similar distributions. When the Si atoms were deposited onto the growth front, InAs QDs have formed. The AFM measurements indicate that the Si doping at the self-limiting growth stage does not affect the growth of the QDs.

Then we measured the conductive current distribution on the surface InAs QDs with different Si doping levels, as shown in Fig. 2. The +3V forward voltage was biased with respect to the grounded GaAs substrate. Several interesting phenomena can be observed from Fig. 2: (I) the Si-doped QDs are more conductive than the undoped ones; (II) The current value decreases quickly from ~0.4 nA to ~0.1 nA when the doping level increases from $1\times10^{11}$ cm$^{-2}$ to $5\times10^{11}$ cm$^{-2}$, as shown by Fig. 2(c) and 2(d); (III) with the increase of Si doping levels, more and more QDs' conductivity becomes better, as shown by Fig. 2(b)-2(d). Inoue *et al.*[3] also observed the phenomenon (I) and attributed it to the rising of Fermi level in the Si-doped QDs. However, this cannot explain the phenomenon (II). Alternatively, we propose another explanation as described below. Since the Si-doped InAs layer is nominally less than 0.24 MLs, the schematic band structure of the surface Si-doped QDs is shown in Fig. 2(e).[7,8] When the positive voltage is applied on the conductive AFM tip, the electrons confined inside the doped InAs layer will be pulled away by the electronic field. The resultant charged InAs layer will draw the electrons from the underlying GaAs layer. In this case, the Si doped InAs layer supplies the extra electrons and transfers the electrons, thus improving the conductivity. This can explain the phenomenon (I). However, as the doping level increases, the Coulomb force and possible scattering between the electrons and the charged Si atoms become strong enough that they reduce the electron transport. As a result, the current value decreases, but does not disappears; this can explain the phenomenon (II). Phenomenon (III) can be explained as follows: when the Si atoms were deposited, some of them migrated into the existed InAs QDs together with In atoms. The more Si atoms that were deposited, the more the InAs QDs were Si doped, and thus their conductivity increased.

Fig. 3(a) depicts the ground-state (GS) PL peak intensities of the Si-doped InAs QDs from 13.7 K to 300 K. At low temperatures (<100 K), the PL intensities of the Si-doped QDs with different levels sort as follows: $I_{1\times10^{11} cm^{-2}} > I_{5\times10^{11} cm^{-2}} >$ $I_{4\times10^{10}cm^{-2}} > I_{undoped} \approx I_{1\times10^{12}cm^{-2}}$. With the increase of temperature, the GS PL intensities of the QDs undoped, doped with $4\times10^{10}cm^{-2}$ and $1\times10^{11}cm^{-2}$ quench similarly, but with different onset temperatures: the former two begin to drop at 150K while the latter one begins to drop at 175K. When doped with $5\times10^{11}cm^{-2}$ and $1\times10^{12}cm^{-2}$, the QDs PL intensities decrease very slowly until 230K and then steeply at higher temperatures. These results indicate that the Si doping can delay the temperature quenching of PL intensity, and that the larger the Si doping density, the higher the onset temperature for quenching. Fig. 3(b) shows the PL spectra of Si-doped InAs QDs at 300 K. The GS PL peak intensity of the InAs QDs doped with $1\times10^{11}$ cm$^{-2}$ and $5\times10^{11}$ cm$^{-2}$ are twenty and thirty-five times stronger than that of the undoped ones, respectively. Moreover, the first excited-state (ES) peaks are obvious when doping density reaches $5\times10^{11}$ cm$^{-2}$ and $1\times10^{12}$ cm$^{-2}$.

The Si atom usually substitutes for the In atom in InAs and the length of the formed Si-As bond is shorter than that of the In-As bond,[5] so the Si-doped InAs layer can relax the strain remaining inside the InAs QDs and thus reduce the non-radiative centers in the InAs/GaAs interface. However, doping too many Si atoms into the InAs QDs can destroy the crystal lattice and reintroduce non-radiative centers again. This can explain the dependence of the PL intensity on doping levels at low temperatures (<100K) in Fig. 3(a). In addition, the relaxed strain inside the InAs QDs can reduce the band gap and thus red shift the PL peak, as indicated by Fig. 3(b). The Si atoms inside the InAs QDs can also deepen the potential well in the InAs QDs.[9] This energy difference is denoted by ΔE in the schematic band structure of the embedded QDs, shown in Fig. 2(f). The value of ΔE should increase with increasing the doping levels. The larger ΔE is, the deeper the potential well is, making it more difficult for carriers to escape from the QDs. This can explain the delayed temperature quenching in Fig. 3(a). In order to demonstrate this point further, as shown in Fig. 3(c), we used the integrated PL intensity and the fitting formula below to derive the thermal activation energy:

$$I(T) = \frac{I_0}{1 + C \exp(-E_A/kT)} \cdot ^{10} \quad (1)$$

The activation energies are 292±10meV for the undoped QDs, 374±21meV for doping of $4\times10^{10}$cm$^{-2}$, 380±11meV for doping of $1\times10^{11}$cm$^{-2}$, 422±31meV for doping of $5\times10^{11}$cm$^{-2}$, and 498±15meV for doping of $1\times10^{12}$cm$^{-2}$. The activation energies of the Si-doped QDs increase with the doping levels, and all are much larger than that of the undoped ones or the theoretical value of 300meV,[11] indicating a deeper potential well after Si doping. Another result of the deepened potential well is that the Fermi level inside QDs rises, making it easier for photo-generated carriers to occupy the first ES. This can explain the first ES PL peak beginning to appear in Fig. 3(b), as the doping level exceeds $5\times10^{11}$cm$^{-2}$.

Fig. 3(d) shows the full width at half maximum (FWHM) of the GS PL intensity. At 300K, except for the QDs doped to $1\times10^{12}$cm$^{-2}$, the FWHM of the PL peak of the Si-doped QDs are all smaller than the FWHM of the undoped QDs. This indicates that optimized Si doping levels can reduce the FWHM of the PL peak. Another phenomenon is that from 175K to 300K, the FWHM variations of the Si-doped QDs are much less than that of the undoped ones. This indicates that the Si doping can prevent carriers from thermally migrating from the smaller QDs to the larger ones.[12]

Different from our proposed model, Inoue *et al.*[3] attributed the PL enhancement of Si-doped InAs QDs to excess electrons inactivating the defect state inside the QDs. However, this can not explain the PL intensity decreasing at <100 K when doping density exceeds $1\times10^{11} cm^{-2}$ as we observe, and in both their work and ours, the obtained activation energies are much larger than the theoretical value of 300meV.[11]

In summary, doping Si atoms into the InAs QDs after the QDs initially forms can red shift the GS PL peaks, and appropriate doping levels can greatly enhance the PL intensity, delay the temperature quenching and reduce the linewidth. These results are attributed to a supposed positive-charged, strain-relaxed Si-doped InAs layer inside the InAs QDs.

This work was funded partly by the National Science Foundation of China (Nos. 60876033, 61076050, 61021003), and partly by the China Postdoctoral Science Foundation (20100470529).

**Figures Captions**

FIG. 1 AFM images of the surface InAs QDs with different Si doping densities. (a) undoped; (b) $1\times10^{11}$cm$^{-2}$; (c) $5\times10^{11}$cm$^{-2}$; (d) $1\times10^{12}$cm$^{-2}$.

FIG. 2 Current distributions of the surface InAs QDs with different Si doping densities: (a) undoped; (b) $4\times10^{10}$cm$^{-2}$; (c) $1\times10^{11}$cm$^{-2}$; (d) $5\times10^{11}$cm$^{-2}$. (e) and (f) are the schematic band structures for the surface InAs QDs and embedded InAs QDs, respectively.

FIG. 3 (a) The GS PL intensity as a function of temperature from 13.7 to 300K. (b) PL spectra from InAs QDs with different Si doping densities, measured at 300K; (c) and (d) are the FWHM of the GS PL peak and the integrated PL intensity as a function of temperature from 13.7 to 300 K, respectively.

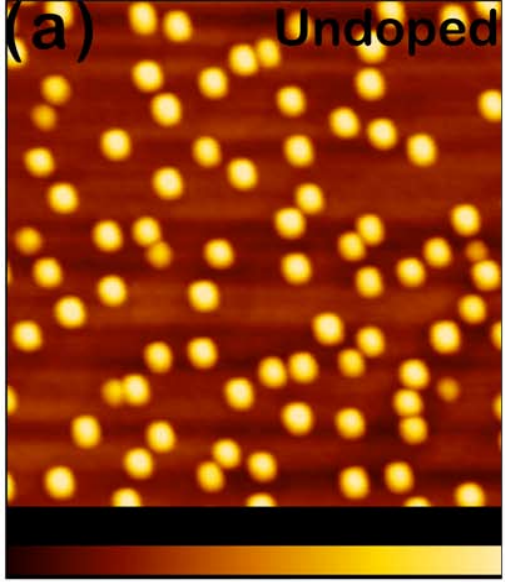 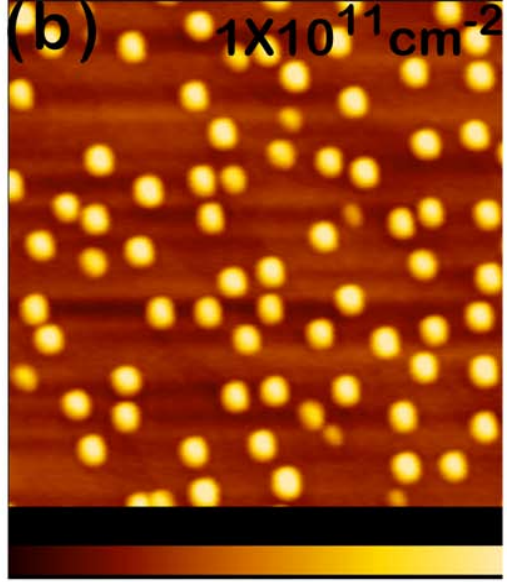
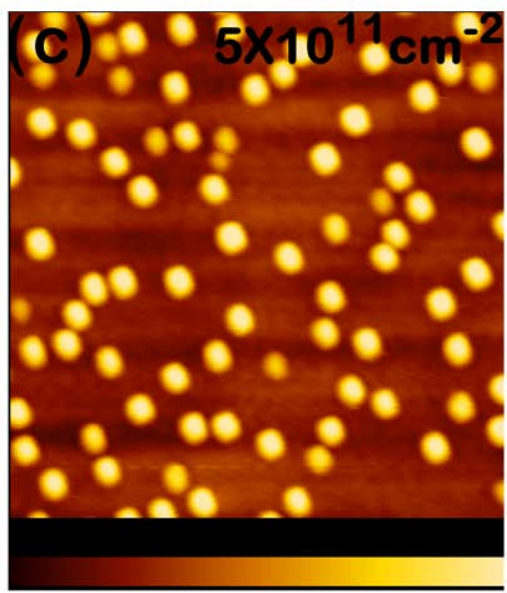 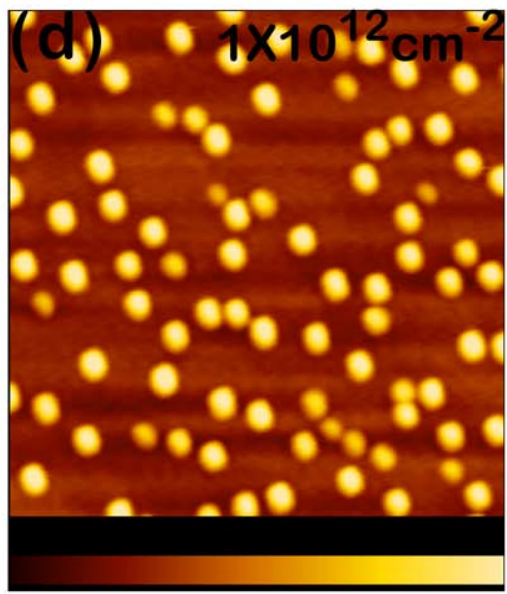

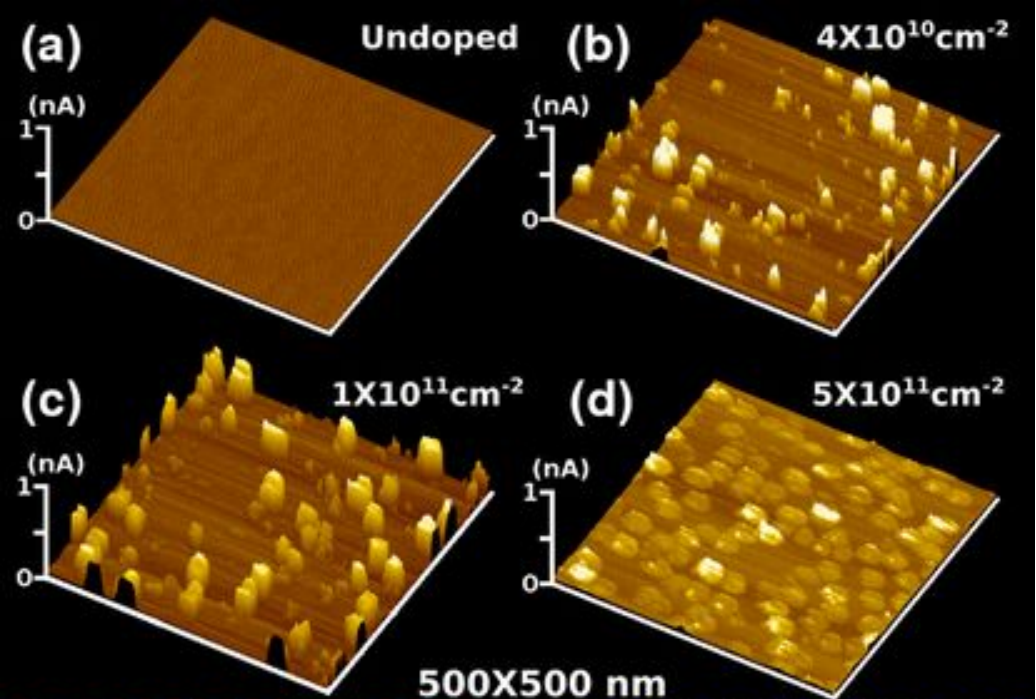

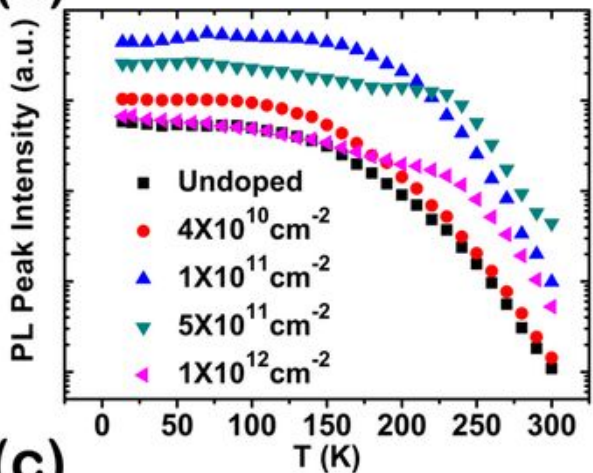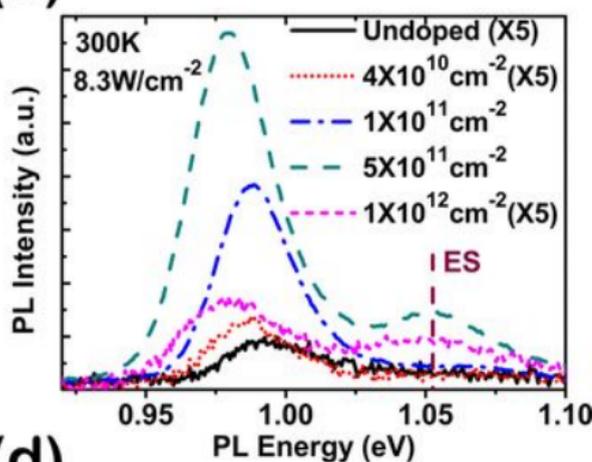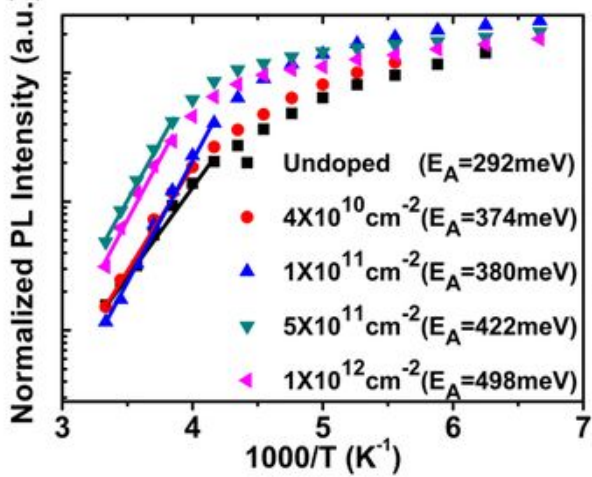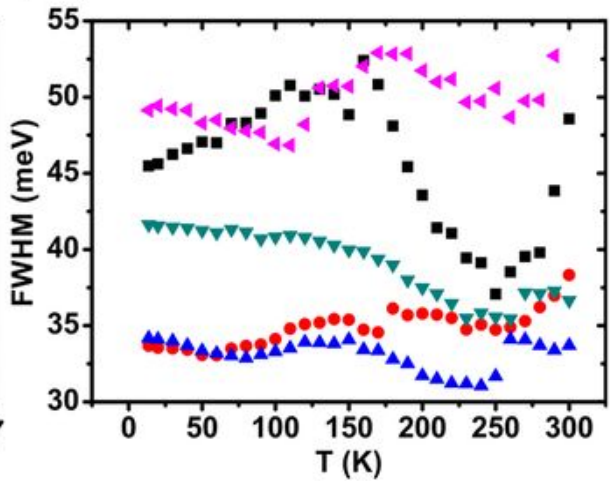